\documentclass[notitlepage,superscriptaddress,nofootinbib,tightenlines,preprintnumbers,twocolumn,floatfix]{revtex4-2}
\usepackage{amsmath,verbatim,latexsym,amssymb,indentfirst,mathrsfs,mathtools,amsthm,bbm,bm,hyperref,url,cancel,subcaption,enumitem,dsfont}
\usepackage[font=small,labelfont=bf,format=plain,justification=raggedright,singlelinecheck=false]{caption}
\usepackage{verbatim,indentfirst}
\usepackage[title,titletoc]{appendix}
\usepackage{color}
\usepackage[dvipsnames]{xcolor}
\usepackage[normalem]{ulem}
\usepackage{tikz}
\usepackage{setspace}
\usetikzlibrary{quantikz}
\def \qmeasurement {unitary sensor}  
\def \qmeasurements{unitary sensors}  
\def \Qmeasurements {Unitary sensors} 

\usepackage{graphicx}
\usepackage{placeins}
\usepackage{dcolumn}
\usepackage{bm}
\usepackage{multirow}

\begin{document}

\preprint{APS/123-QED}

\title{
Information gain and measurement disturbance for quantum agents}

\author{Arthur O. T. Pang}
\email{arthur.pang@mail.utoronto.ca}
\affiliation{Department of Physics and Centre for Quantum Information \& Quantum Control, University of Toronto$,$ Toronto$,$ Ontario$,$ Canada}
\author{Noah Lupu-Gladstein}
\email{nlupugla@uottawa.ca}
\affiliation{National Research Council of Canada, Ottawa, Ontario, Canada}
\affiliation{Department of Physics, University of Ottawa, Ottawa, Ontario, Canada}
\author{Y. Batuhan Yilmaz}
\email{ybylmaz@physics.utoronto.ca.}
\affiliation{Department of Electrical and Computer Engineering, University of Toronto$,$ Toronto$,$ Ontario$,$ Canada}
\affiliation{IonQ, Inc. Toronto, Ontario, Canada}
\author{C. Pria Dobney}
\email{pria.dobney@mail.utoronto.ca}
\author{Rui Jie Tang}
\email{ruijie.tang@mail.utoronto.ca}
\affiliation{Department of Physics and Centre for Quantum Information \& Quantum Control, University of Toronto$,$ Toronto$,$ Ontario$,$ Canada}

\author{Aharon Brodutch}
\email{brodutch@ionq.co}
\affiliation{IonQ, Inc. Toronto, Ontario, Canada}

\author{Aephraim M. Steinberg}
\email{steinberg@physics.utoronto.ca}
\affiliation{Department of Physics and Centre for Quantum Information \& Quantum Control, University of Toronto$,$ Toronto$,$ Ontario$,$ Canada}
\affiliation{Canadian Institute for Advanced Research$,$ Toronto$,$ Ontario$,$ Canada}
\date{$8^{\mathrm{th}}$ April 2026}

\begin{abstract}
The traditional formalism of quantum measurement (hereafter ``TQM'')  describes processes where some properties of quantum states are extracted and stored as classical information. While TQM is a natural and appropriate description of how {\em humans} interact with quantum systems, it is silent on the question of how a more general, {\em quantum}, agent would do so.  How do we describe the observation of a system by an observer with the ability to store not only classical information but quantum states in its memory? 
In this paper, we extend the idea of measurement to a more general class of  sensors for quantum agents which interact with a system in such a way that the agent's memory stores information (classical or quantum) about the system under study.  For appropriate sensory interactions, the quantum agent may ``learn'' more about the system than would be possible under any set of classical measurements -- but as we show, this comes at the cost of additional measurement disturbance.  We experimentally demonstrate such a system and characterize the tradeoffs by considering the channel capacity required to erase the effect of a measurement.
\end{abstract}

\maketitle

\section{Introduction}
The traditional quantum measurement formalism (which we will term ``TQM") models an interaction between a quantum system and a sensor, under the implicit assumption that a sensor is in a sense classical: a device which interacts with a system and generates classical information as a result of this measurement.
Understanding and modeling measurement has of course been of central importance to the development and interpretation of quantum theory.
It is natural to consider the power and limitations of sensors which go beyond this classical framework (we use the term ``sensor'' instead of ``measurement'' to avoid confusion between the already fraught topic of quantum measurement theory and the extension we wish to discuss \cite{sheeple}).
After all, even humans (so far as we know) are governed by the laws of quantum physics, and there are even speculations that the latter are essential in the operation of our brains \cite{HAMEROFF201439}.  Moreover, with the rapid advances in artificial intelligence and quantum computing, it is conceivable that a ``quantum agent'' (e.g., some sort of robot controlled by a quantum computer) with the ability to probe the quantum world may exist in a not-too-distant future.
Such an agent would store the quantum information gained from their interactions in the form of qubits in quantum memories. 
The notion of quantum agency has been discussed in other works such as \cite{quantum_adaptive_agent,agent_learning,quantum_adaptive_agents,sheeple}, and most notably in the Wigner's friend paradox \cite{Wigner_bruker,Wigner_wiseman,Frauchiger2018}, but the capabilities and properties of the sensory interaction were never the focus of these works. What form would this agent's observations, stored in its quantum memory, take?  What about the back-action of its sensory interactions on the system being studied?  We explore these questions in this work.


Upgrading an agent's memory from classical to quantum endows it with fundamentally new options for learning about a system.
A classical agent is restricted to extracting partial information as classical bits through projective measurement, and must use TQM to convert quantum states into classical data. This approach is inherently limited. For a spin-1/2 qubit system, for instance, a classical agent can measure a single Pauli component, 
$\sigma_i, i\in \{x,y,z\}$, using an interaction like $H_{\text{int}}\propto
\sigma_{i}^{\text{system}}\otimes\sigma_{i}^{\text{probe}}$. 
While this yields a precise estimate of that one component, the measurement inevitably disturbs the non-commuting observables, completely destroying information about the other two Pauli bases.

On the other hand, a quantum agent, equipped with a quantum memory, can in principle learn the entire system state by acquiring the state itself.
This requires an interaction that transfers the state from the system to the agent's memory, such as a SWAP operation, which can be represented by the following unitary:
\begin{multline}\label{eq:swap_full}
    \hat{\text{SWAP}}=\\
    \left(\hat{\mathds{I}}\otimes\hat{\mathds{I}}
    +\hat\sigma_x\otimes\hat\sigma_x
    +\hat\sigma_y\otimes\hat\sigma_y
    +\hat\sigma_z\otimes\hat\sigma_z\right)/2,
\end{multline}

A SWAP interaction allows the agent to gain complete information about all Pauli components at once. However, the no-cloning theorem dictates that this total information gain comes at the highest cost: the original system's state is completely overwritten—a maximal disturbance.

In this paper, we consider the simplest case, of two-dimensional (qubit) systems and unitary sensors.
We focus our discussion of qubit-qubit unitaries in the context of \qmeasurements{} for quantum agents.
A \qmeasurement{} is a unitary interaction that generates correlations between an initial quantum state of a system of interest $\mathcal{S}$ and the post-interaction state of some memory quantum system $\mathcal{M}$. 
We use quantum mutual information as a quantifier of the correlations generated and back-action imparted by these unitaries (section \ref{sec:QMI}).
To further understand the back-action imparted by a \qmeasurement{} and the information gain, we will use the operator-Schmidt decomposition (section \ref{sec:osdecomp}) to quantify the resources needed to restore the system $\mathcal{S}$ to its initial state $\ket{\psi}$ and undo the effect of the \qmeasurement{} (section \ref{sec:erase}).


Since the back-action caused by a full SWAP unitary exceeds the disturbance induced by any measurement we perform through TQM, a natural question arises: if we perform weaker quantum measurements by implementing partial SWAP unitaries, which still leave the system $S$ partially correlated with its initial state $|\psi\rangle$, up to a point where this weak quantum measurement disturbance is quantitatively the same as the disturbance caused by the TQM, would TQM and the weak quantum measurement scheme be fundamentally different? The answer is yes.
We have discovered that regardless of how weak our quantum measurement or classical measurement may be, the amount of resources required to undo the measurement action are fundamentally dependent on the Schmidt rank of the unitary sensor.
These two measurement schemes have different Schmidt ranks: any classical sensor is of rank 1, while quantum sensors can be of higher ranks.
We experimentally implement TQM and SWAP-like interactions to empirically demonstrate the qualitative differences between sensor unitaries of different rank.

\section{Measurement as a Unitary}\label{sec:measurement_task}
\subsection{Traditional Quantum Measurement Formalism}\label{sec:TQMform}
Traditionally, quantum measurement has been formulated as a tool to translate the properties of a quantum state into probabilistic classical events, and although TQM is meant to have results given by classical events, John von Neumann provided \footnote{There are other TQM formalisms, such as the PVM and POVM formalisms. For the purposes of this work, von Neumann's scheme represents a sufficient representation of TQM.} a fully quantum treatment of quantum measurements \cite{vNeumann}.
In the von Neumann measurement scheme, a measurement device with a quantum memory ($\mathcal{M}$), whose initial wavefunction is well-localized, can measure an observable $\hat{A}^\mathcal{S}$ on an unknown system of interest ($\mathcal{S}$) by implementing the Hamiltonian
\begin{equation}
    \hat{H}_{vN}=g\hat{A}^\mathcal{S}\otimes\hat{p}^\mathcal{M},
\end{equation}
where $\hat{p}^\mathcal{M}$ is the momentum operator acting on the quantum memory. Implementing the Hamiltonian for a period of time, an unknown initial system state $\ket{\psi}^\mathcal{S}=\sum c_i\ket{A_i}^\mathcal{S}$ becomes entangled with the memory state, resulting in the joint state $\ket{f}^\mathcal{M,S}~=~\sum c_i\ket{A_i}^\mathcal{S}~\otimes~\ket{x=g\cdot~t\cdot ~A_i}^\mathcal{M}$. 

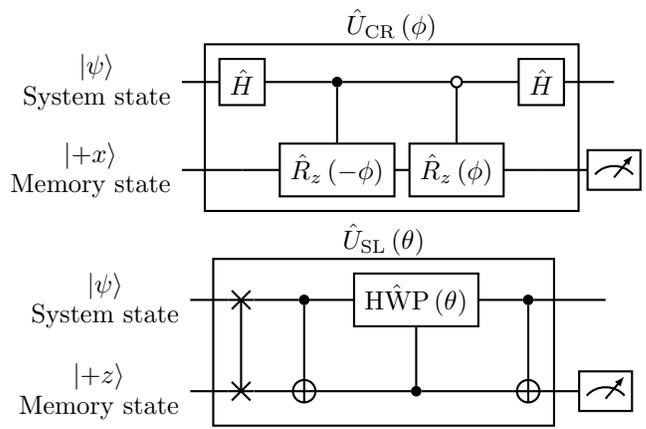
\begin{figure}
    \centering
    \begin{quantikz}
        \lstick{$\ket{\psi}$\\System state} & \gate{\hat{H}}\gategroup[wires=2,steps=4,style={inner sep=2pt}]{ $\hat{U}_{\text{CR}}\left(\phi\right)$} & [-0.3cm]\ctrl{1} & [-0.3cm]\octrl{1} & [-0.3cm]\gate{\hat{H}} & [-0.2cm]\qw \\
        \lstick{$\ket{+x}$\\Memory state} & \qw & \gate{\hat{R}_z\left( - \phi\right)} & \gate{\hat{R}_z\left( \phi\right)} & \qw & \meter{}
    \end{quantikz}

    \begin{quantikz}
        \lstick{$\ket{\psi}$\\System state} & \swap{1}\gategroup[wires=2,steps=4,style={inner sep=2pt}]{ $\hat{U}_{\text{SL}}\left(\theta\right)$} & \ctrl{1} & \gate{\hat{\text{HWP}}\left(\theta\right)} & \ctrl{1} & \qw \\
        \lstick{$\ket{+z}$\\Memory state} & \targX{} & \targ{} & \ctrl{-1} & \targ{} & \meter{}
    \end{quantikz}
    \caption{Top: Circuit diagram of the controlled-rotation unitary $\hat{U}_{\text{CR}}\left(\phi\right)$ (equation~\ref{eq:cpauli_unitary}), with its physical setup presented in figure \ref{fig:cpauli_setup}.
    The Hadamard gates on the system state correspond to the Mach-Zehnder interferometer opened with a polarizing beam splitter and closed with a 50:50 beamsplitter, and the controlled z-rotations correspond to the liquid crystal waveplate (LCWP) in each arm of the Mach-Zehnder interferometer.\\Bottom: Circuit diagram of SWAP-like unitary $\hat{U}_{\text{SL}}\left(\theta\right)$ (equation~\ref{eq:SWAP_measurement_unitary}), with its physical setup presented in figure \ref{fig:swap_setup}.  $\hat{U}_{\text{SL}}\left(\theta\right)$ consists of a SWAP gate due to our preparation procedures; a CNOT gate from the polarizing beamsplitter (PBS) where the first (second) interferometer closes (opens); a controlled half-waveplate (HWP) rotation from the HWP in the second Sagnac interferometer; and another CNOT gate from the PBS where the second (third) interferometer closes (opens).
    The unitary performed by the HWP is given by $\hat{\text{HWP}}\left(\theta\right)=-\left(\cos(\theta)\hat{\sigma}_z + \sin(\theta)\hat{\sigma}_x\right)$.}
    \label{fig:unitary_circuits}
\end{figure}

We will use the controlled-rotation (CR) unitary to represent TQM.
The \qmeasurement{} is given by
\begin{multline}\label{eq:cpauli_unitary}
\hat{U}_{\text{CR}}\left(\phi\right)=\\
\ket{+x}\!\!\bra{+x}^\mathcal{S}\otimes\hat{R}_z\left(\phi\right)^\mathcal{M}+\ket{-x}\!\!\bra{-x}^\mathcal{S}\otimes\hat{R}_z\left(-\phi\right)^\mathcal{M},
\end{multline} 
This can be expressed in matrix form in the \textit{z}-basis as:
\begin{align}
   \hat{U}_{\text{CR}}(\phi) = 
\begin{pmatrix}
\cos(\frac{\phi}{2}) & 0 & -i\sin(\frac{\phi}{2}) & 0 \\
0 & \cos(\frac{\phi}{2}) & 0 & i\sin(\frac{\phi}{2}) \\
-i\sin(\frac{\phi}{2}) & 0 & \cos(\frac{\phi}{2}) & 0 \\
0 & i\sin(\frac{\phi}{2}) & 0 & \cos(\frac{\phi}{2})
\end{pmatrix} 
\end{align}
where $\hat{\mathds{I}}$ is the identity operator and $\hat{R}_z$ is a rotation along the z-axis in the Bloch sphere, with its maximum interaction strength at $\phi=\pi/2$. This \qmeasurement{} is generated by the von Neumann Hamiltonian with the observable $\hat{A}^\mathcal{S}=\hat{\sigma}_x$, quantum memory momentum operator of $\hat{p}^{\mathcal{M}}=\hat{\sigma}_z$, and with an interaction time of $t=\phi/g$. Under this \qmeasurement{}, we can measure $\hat{\sigma}_x$ by preparing the memory as one of the eigenstates of $\hat{\sigma}_x$ and subsequently projecting it onto an eigenstate of $\hat{\sigma}_y$ after the interaction. The corresponding circuit diagram of this \qmeasurement{} can be found in figure \ref{fig:unitary_circuits}.

von Neumann unitaries extract information about the observable $\hat{A}^\mathcal{S}$ but reveal nothing about complimentary observables.
For the \qmeasurement{} $\hat{U}_{\text{CR}}\left(\phi\right)$, this translates to the fact that it can only facilitate the measurement of $\hat{\sigma}_x$, but not $\hat{\sigma}_y$ or $\hat{\sigma}_z$. 


\begin{figure}[htpb]
    \centering
    \includegraphics[width=0.95\columnwidth]{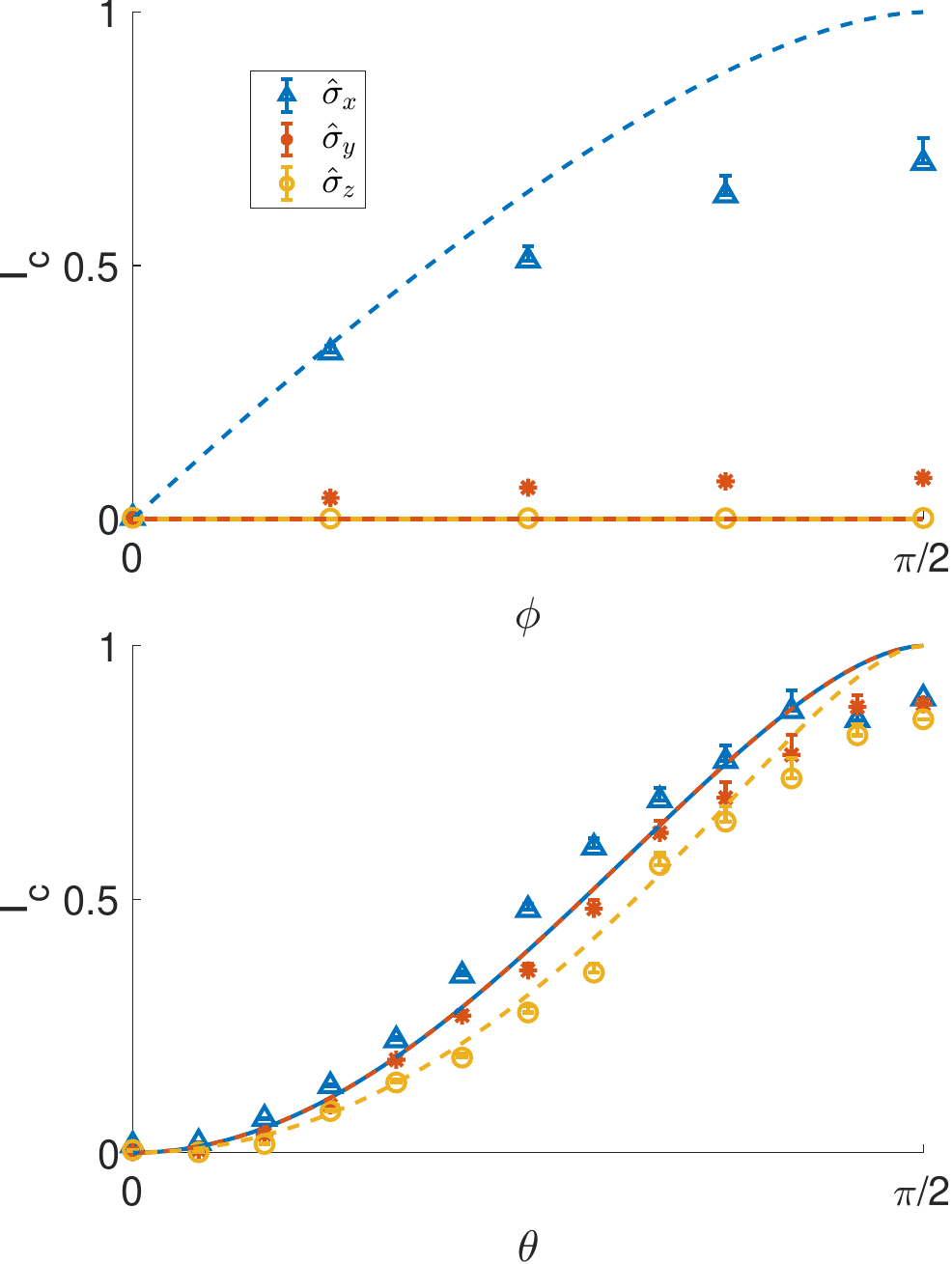}
    \caption{Plots of the classical mutual information ($I_C$, defined in Equation~(\ref{eq:CMI})) between Pauli measurements on the initial system and post-interaction memory in the labeled basis.
    Here, points represent experimental values and dashed lines represent the theoretical $I_c$ values calculated from the intended unitaries.
    These plots represents how well the agent can sense initial system states along $\hat\sigma_x$, $\hat\sigma_y$, and $\hat\sigma_z$ eigen-bases.
    The top plot represents $\hat{U}_{\text{CR}}\left(\phi\right)$ and the bottom plot represents $\hat{U}_{\text{SL}}\left(\theta\right)$.
    The horizontal axis of both plots corresponds to the (arbitrary) value of tuning parameters $\phi$ and $\theta$ for the two unitaries given in equations \ref{eq:cpauli_unitary} and \ref{eq:SWAP_measurement_unitary}, with the left corresponding to the identity operator and the right being the unitaries at their maximum interaction strength.
    The agent never attains information about the $\hat\sigma_y$ and $\hat\sigma_z$ eigen-bases using $\hat{U}_{\text{CR}}\left(\phi\right)$, whereas the information for all three bases increases for $\hat{U}_{\text{SL}}\left(\theta\right)$ as the maximum interaction strength is approached.
    Due to experimental imperfections, $I_c$ for $\hat\sigma_x$ measurement for  $\hat{U}_{\text{CR}}\left(\phi\right)$ and all three measurement bases for  $\hat{U}_{\text{SL}}\left(\theta\right)$ does not reach the maximum value of 1. Here, the plotted error bars represent systematic error caused by phase randomization of our interferometer. These systematic errors are asymmetric, as adding randomness to any system only decreases the amount of mutual information. Errors due to imprecise waveplate rotation and off-axis LCWP rotations are not included in the error bars. Appendix \ref{secap:Errors} describes these errors in detail.}
    \label{fig:CMI}
\end{figure}

\subsection{\Qmeasurements{} for Quantum Agents}\label{sec:Quesure}


The SWAP unitary (equation \ref{eq:swap_full}), unlike von Neumann unitaries, encodes all statistics of any observables of the quantum state of interest $\ket{\psi}$ in the post-interaction memory state.
Experimentally (as will be further discussed in section \ref{sec:exp_setup}), we implement $\hat{U}_{\text{SL}}\left(\theta\right)$ to represent the class of bipartite qubit-unitaries that are SWAP-like
\begin{multline}\label{eq:SWAP_measurement_unitary}
    \hat{U}_{SL}(\theta) = \frac{1}{2} [
    -(1+\sin (\theta-\pi/2))\,\mathds{I}^\mathcal{S}\otimes\mathds{I}^\mathcal{M} -\\ 
    (1-\sin (\theta-\pi/2))\hat{\sigma}_z^\mathcal{S}\otimes\hat{\sigma}_z^\mathcal{M} + \\
    i\,\cos (\theta-\pi/2) \hat{\sigma}_x^\mathcal{S}\otimes\hat{\sigma}_y^\mathcal{M} - 
    i\,\cos (\theta-\pi/2) \hat{\sigma}_y^\mathcal{S}\otimes\hat{\sigma}_x^\mathcal{M}]
\end{multline}

This can be expressed in matrix form as:
\begin{align}
    \hat{U}_{SL}(\theta) &= \begin{pmatrix}
       1 & 0 & 0 & 0\\
       0 & - \cos(\theta) & -\sin(\theta) & 0\\
       0 & - \sin(\theta) & \cos(\theta) & 0\\
       0 & 0 & 0 & 1
    \end{pmatrix}
\end{align}
where $\theta$ is a tuning parameter that takes our \qmeasurement{} from identity at $\theta = 0$ to its maximum interaction strength at $\theta = \pi/2$. $\hat{U}_{\text{SL}}\left(\theta\right)$ at its maximum interaction strength is equivalent to the SWAP unitary followed by a controlled-phase gate. We choose to employ $\hat{U}_{\text{SL}}\left(\theta\right)$ due to the ease of implementing it experimentally, and for the intended analysis of this paper, the SWAP and SWAP-like unitary  $\hat{U}_{\text{SL}}\left(\theta\right)$ both serve the same purpose. For  $\hat{U}_{\text{SL}}\left(\theta=\pi/2\right)$, an initial memory state of $\ket{+z}$ would result in the memory state after the interaction being the initial system state $\ket{\psi}$. The corresponding circuit diagram for $\hat{U}_{\text{SL}}\left(\theta\right)$ can be found in figure \ref{fig:unitary_circuits}.

Figure \ref{fig:CMI} plots the classical mutual information $I_c$ between the results of Pauli measurements on the initial system state and post-interaction memory state in the $\hat\sigma_x$, $\hat\sigma_y$, and $\hat\sigma_z$ eigen-bases for $\hat{U}_{\text{CR}}$ and $\hat{U}_{\text{SL}}$, where
\begin{multline}\label{eq:CMI}
    I_c\left(X;Y\right)=H_s(Y)-H_s(Y|X)\\=H_s(X)+H_s(Y)-H_s(X,Y).
    \end{multline}
Here $H_s(\cdot)$ is the Shannon entropy, $X$ is the set of initial system states, and $Y$ being the projections of the post-interaction memory state.
Unlike von Neumann-type unitaries,  $\hat{U}_{\text{SL}}\left(\theta\right)$ generates correlations between the initial system state and post-interaction memory state on all three Pauli bases. It is important to note that although correlations for all three Pauli bases can be stored in the quantum memory due to the \qmeasurement{}, they are not all simultaneously extractable as classical information due to the uncertainty principle given in TQM.
On the other hand, an agent interested in processing states quantum mechanically will find its memory state to be a more complete representation of the state $\psi$ by using the sensor  $\hat{U}_{\text{SL}}\left(\theta=\pi/2\right)$.

\section{Experimental setup}\label{sec:exp_setup}

Both of our experimental setups are built with free-space optics in conjunction with a spontaneous parametric down-conversion (SPDC) photon source using periodically poled potassium titanyl phosphate (PPKTP) resulting in type-II co-linear down-conversion that generates a signal and an idler photon. The source gives $\sim 63,000$ heralded single photons per second. After the single-mode fibre coupling into a detector, we detect $\sim 8,000$  and $\sim 1,000$ photon pairs per second for the $\hat{U}_{\text{CR}}\left(\phi\right)$ and $\hat{U}_{\text{SL}}\left(\theta\right)$ setups correspondingly, with losses due to unwanted absorption, reflection, and single-mode coupling inefficiencies.

\begin{figure}
    \centering
    \includegraphics[width=0.75\columnwidth]{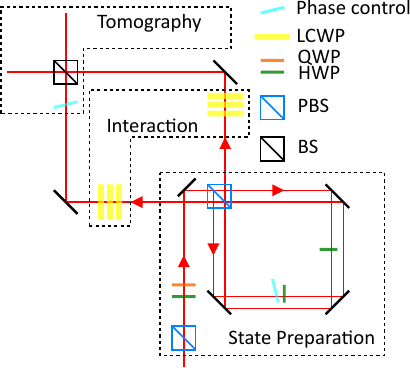}
    \caption{Experimental setup for the agent using sensor $\hat{U}_{\text{CR}}\left(\phi\right)$, defined in equation~\ref{eq:cpauli_unitary}. Signal photons first go through a Sagnac interferometer with a polarizing beam splitter (PBS) and an associated set of waveplates such that the system state is encoded in the path degree of freedom and the memory state is encoded in the polarization degree of freedom. A Mach-Zehnder interferometer follows, where the \qmeasurement{} is facilitated by the LCWPs placed in the two arms of the interferometer. The Mach-Zehnder is closed with a non-polarizing beam splitter (BS), after which polarization tomography takes place.}
    \label{fig:cpauli_setup}
\end{figure}

The experimental setup for the agent using sensor $\hat{U}_{\text{CR}}\left(\phi\right)$ consists of a state preparation stage in a Sagnac interferometer followed by the \qmeasurement{} of $\hat{U}_{\text{CR}}\left(\phi\right)$ facilitated by liquid crystal waveplates (LCWPs) in a Mach-Zehnder interferometer. In this experiment, the quantum memory corresponds to the signal photon polarization and the system corresponds to the signal photon path in the Mach-Zehnder interferometer. By tuning the LCWPs optical activity, the unitary given in equation (\ref{eq:cpauli_unitary}) can be implemented for arbitrary $\phi$ values. In this experiment, the signal photon is passed into the setup and the herald is used as a triggering mechanism for our detectors. The corresponding circuit and setup diagram of $\hat{U}_{\text{CR}}\left(\phi\right)$ can be found in figures \ref{fig:unitary_circuits} and \ref{fig:cpauli_setup}.

Through the Sagnac interferometer state-preparation stage, we prepare a polarization-path product state: {$(a|\rightarrow\rangle+b|\downarrow\rangle)_{\text{sys}}\otimes |+x\rangle_{\text{mem}}$} with one photon.
Initially, the qubit passing through the first PBS (at the bottom of the diagram in figure~\ref{fig:cpauli_setup}) is prepared in the state $|H\rangle\otimes |\uparrow\rangle$, where arrows represent the qubit's direction of travel in the diagram in figure~\ref{fig:cpauli_setup}.
Subsequently, a quarter-waveplate (QWP) and HWP pair transform the system's state into $(a|H\rangle+b|V\rangle)_{\text{sys}}$ while the path qubit remains in the state $|\uparrow\rangle_{\text{mem}}$.
Note that the Sagnac interferometer will be used to prepare the memory qubit into $|+x\rangle$ and swap its state with that of the system.
The second PBS is placed at the opening of the Sagnac interferometer, in the center of the diagram in figure~\ref{fig:cpauli_setup}.
The PBS transmits horizontal light and reflects vertical light, resulting in the state $a|\rightarrow H\rangle+b|\downarrow V\rangle$ after the PBS.
Placed after the transmitted port of the PBS is a HWP rotated to $22.5^\circ$ which transforms $|H\rangle$ into $|+x\rangle$. 
Similarly, another HWP rotated to $-22.5^\circ$ is placed after the reflected port to transform $|V\rangle$ into $|+x\rangle$. This configuration yields the state $(a|\rightarrow\rangle+b|\downarrow\rangle)\otimes|+x\rangle$. These two paths then recombine through the same PBS that opened the interferometer, preparing the desired product state, $(a|\rightarrow\rangle+b|\downarrow\rangle)_{\text{sys}}\otimes |+x\rangle_{\text{mem}}$.
Following the state preparation stage, we incorporate LCWPs set to different $z$-rotation angles, $\hat{R}_z(\pm\phi)$, along the two paths of second interferometer in Mach-Zehnder configuration. These LCWPs act as path-controlled rotation gates on the memory qubit in $|+x\rangle$ state. Finally, closing this second interferometer with a 50:50 BS acts as a Hadamard gate for the system state's path qubit. The memory qubit is projected onto $|\pm y\rangle$ using a polarization tomography setup to extract information about the observable $\hat{\sigma}_z$.

\begin{figure}
    \centering
    \includegraphics[width=0.95\columnwidth]{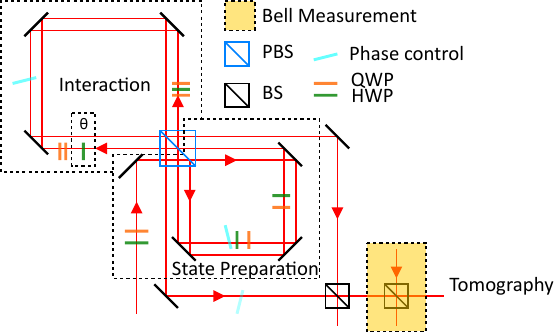}
    \caption{Experimental setup for the agent using sensor $\hat{U}_{\text{SL}}\left(\theta\right)$.  The quantum memory and the system of interest are also the polarization and path degrees of freedom, respectively. The signal photon is injected into a series of three interferometers. The first interferometer is a Sagnac interferometer that opens and closes at a PBS, where it contains a set of waveplates allowing for the preparation of the memory state in the path degree of freedom and the system state in the polarization degree of freedom. This is followed by a second Sagnac interferometer, where the interaction between the path and polarization degrees of freedom takes place. By tuning the angle of the HWP in one of the arms of the interferometer, the \qmeasurement{} can be tuned according to equation (\ref{eq:SWAP_measurement_unitary}). The setup ends with a tomography stage for the path and polarization qubit. To perform erasure for the \qmeasurement{}, a BS and the idler photon may be inserted (highlighted in orange) to perform a polarization Bell state projection.}
    \label{fig:swap_setup}
\end{figure}

The $\hat{U}_{\text{SL}}\left(\theta\right)$ setup consists of a series of three cascaded interferometers, and can be seen in figure~\ref{fig:swap_setup}.
In this experiment, the quantum memory and the system of interest are also the polarization and path degrees of freedom respectively. However, in our state preparation Sagnac interferometer, the system state is prepared in the polarization degree of freedom and the agent's memory in the path degree of freedom.
We initially prepare our input photon in the state $|H\rangle_{\text{mem}}$ which is then transmitted through the PBS.
Following transmission, a HWP sets the photon into the state $|\rightarrow\rangle_{\text{mem}}\otimes(a|H\rangle+b|V\rangle)_{\text{sys}}$. Notably, we have re-labeled the memory state as the path qubit and the system state as the polarization qubit.
This preparation, along with relabeling, can be thought of as a SWAP gate. The prepared qubit subsequently enters the PBS again, which implements a CNOT where the polarization is the control qubit and the path is the target.
This interaction reflects vertically polarized photons to a different path, resulting in the entanglement between the path and polarization states, yielding the state $a|\uparrow H\rangle+ b|\leftarrow V\rangle$.
A HWP positioned along one of the photon paths within the second interferometer functions as a path-controlled HWP on the polarization system state. This HWP at maximum interaction strength is set to $45^\circ$, flipping $|H\rangle$ to $|V\rangle$.
Finally, the closure of the second interferometer using the same PBS acts as the final polarization CNOT gate. In the case of maximum interaction, this disentangles the state into $|V\rangle_{sys}(a|\uparrow\rangle+b|\leftarrow\rangle)_{mem}$. The third interferometer, closed with a 50:50 BS, effectively acts as a Hadamard gate on the path qubit, enabling the measurement of the memory state.

\FloatBarrier

\section{Quantifying information gained and back-action}
\subsection{Quantum Mutual Information}\label{sec:QMI}
A parameter that captures the information gained and total back-action due to any particular \qmeasurement{} is the quantum mutual information (QMI) \cite{sheeple}. 
To measure the information gained and back-action of a unitary sensor, we must quantify the correlations between the past (pre-interaction) system state and future (post-interaction) memory state. Quantum mutual information, quantifies bi-partite correlations at a given time, but not across time.
To measure the correlation between the past and future, we introduce a system ancilla, $\mathcal{S_A}$, that purifies the initial system, $\mathcal{S}$.
For an unknown, and hence maximally mixed input state, the purification is the maximally entangled state, $\ket{\Phi^+}\!\!\bra{\Phi^+}^\mathcal{S,S_A}$.
After the unitary sensor acts on the system and memory, the unchanged ancilla retains an imprint of the system's past.
We define the acquired information of the unitary sensor as the quantum mutual information between the ancilla (i.e. system past) and memory post-interaction.
QMI gives a single parameter quantifier of quantum correlations, which for a \qmeasurement{} $\hat{U}^\mathcal{M,S}$ is given by $I_q^\mathcal{M, S_A}$, which we define as the \textit{acquired information}, where

\begin{equation}\label{eq:QMI_equation_acquired}
    I_q^\mathcal{M, S_A} = H\left(\rho^\mathcal{S_A}\right)
    + H\left(\rho^\mathcal{M}\right)
    - H\left(\rho^\mathcal{M,S_A}\right),
\end{equation}
and
\begin{multline}\label{eq:after_u_state_sys_ancilla}
    \rho^\mathcal{M,S,S_A} = \left(\hat{U}^\mathcal{M,S}\otimes\mathds{I}^\mathcal{S_A}\right)\cdot\\
    \ket{\phi}\!\!\bra{\phi}^\mathcal{M}\otimes\ket{\Phi^+}\!\!\bra{\Phi^+}^\mathcal{S,S_A}\cdot\left(\hat{U}^\mathcal{M,S}\otimes\mathds{I}^\mathcal{S_A}\right)^\dagger
\end{multline}
is the post-interaction density matrix of the combined memory {$\mathcal{M}$}, system {$\mathcal{S}$}, and system ancilla {$\mathcal{S_{A}}$}, with $\rho^\mathcal{M,S_A}$, $\rho^\mathcal{M}$, and $\rho^\mathcal{S_A}$ being the marginal states obtained by tracing $\rho^\mathcal{M,S,S_A}$  over the relevant subspaces.
Here, $H\left(\cdot\right)$ is the von Neumann entropy. $\rho^\mathcal{M,S_A}$ is also the dual-state for the quantum channel that maps system state inputs to memory state outputs.
Finally, $\ket{\phi}\!\!\bra{\phi}^\mathcal{M}$ is the initial memory state.

\begin{figure}
    \centering
    \includegraphics[width=0.95\columnwidth]{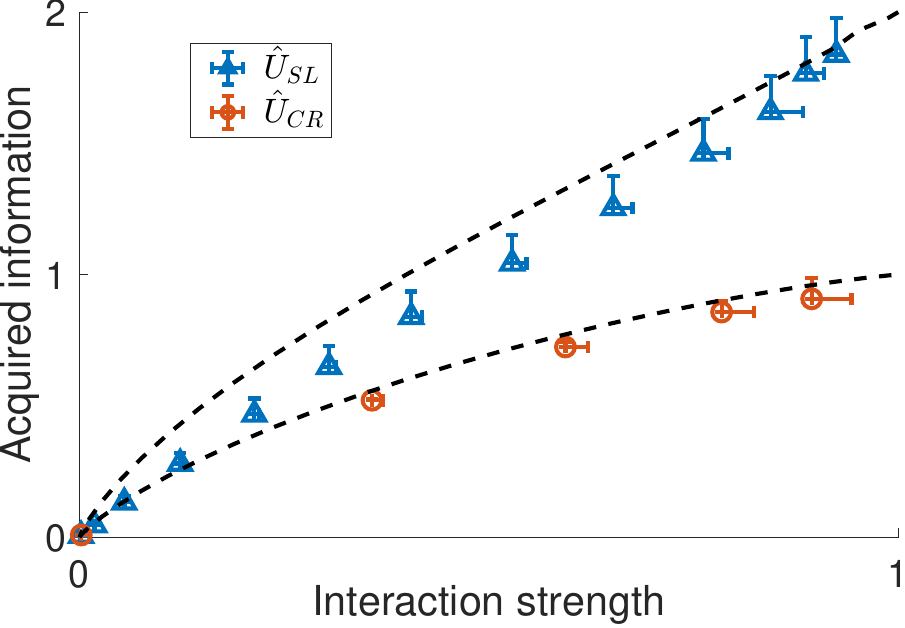}
    \caption{The plot of the acquired information $I_q^\mathcal{M,S_A}$ vs interaction strength represented by maximized classical information $I_{c,max}$ (see equation~\ref{eq:ICmaxXY}) for each of $\hat{U}_{\text{CR}}$ and $\hat{U}_{\text{SL}}$; see equations~\ref{eq:cpauli_unitary} and~\ref{eq:SWAP_measurement_unitary}, respectively. $I_{c,max}$ is calculated from channel tomography by maximizing its value over orthogonal states for encoding and decoding $X,Y$ as described in equation (\ref{eq:CMI}). Here, we observe that the acquired information for $\hat{U}_{\text{SL}}$ is always greater than $\hat{U}_{\text{CR}}$ for a given value of maximized $I_c$. The top and bottom dotted line correspond to the theoretical curve for $\hat{U}_{\text{SL}}$ and $\hat{U}_{\text{CR}}$ respectively. Deviation of mutual information values $\hat{U}_{\text{SL}}$ from their theory line is primarily due to the phase instability in the Mach-Zehnder interferometer, which decreases the value of the acquired information, with a more substantial effect imparted when $\hat{U}_{\text{SL}}$ is close to the identity. Deviation of mutual information values for $\hat{U}_{\text{CR}}$ from their respective theory lines is due to systematic errors causing the presence of correlations between the initial system state in bases other than only in the intended $\hat{\sigma}_x$ basis, resulting in a decrease in the value of $I_{c,max}$. These errors are systematic in nature and their effect is plotted as asymmetric error bars in the plot. The physical nature of these errors is described in appendix \ref{secap:Errors}.}
    \label{fig:QMI_vs_CMI}
\end{figure}

Figure \ref{fig:QMI_vs_CMI} plots the acquired information for $\hat{U}_{\text{CR}}$ and $\hat{U}_{\text{SL}}$ against maximized classical mutual information, $I_{c,max}$, defined in equation~(\ref{eq:ICmaxXY}) below:

\begin{equation} \label{eq:ICmaxXY}
    I_{c,max} = \max_{Y} I_c(X).
\end{equation}
$I_{c,max}$ is calculated via numerical optimization using the estimated channel matrix from channel tomography, by maximizing its value over projective Pauli measurements on the memory $Y$ given a choice of input $X$ as described in equation (\ref{eq:CMI}) and can be interpreted as a quantifier of interaction strength. For the von Neumann unitary $\hat{U}_{\text{CR}}$, the acquired information is bounded by 1, which is due to the fact that von Neumann unitaries are designed to generate classical correlation, which does not exceed 1. For the unitary $\hat{U}_{\text{SL}}$, the acquired information is greater than $I_c$, and in theory should increase to the maximum value of 2 for $\hat{U}_{\text{SL}}$, indicating quantum coherence between the initial system state and the output.

The cost of acquiring information is back-action. In TQM, back-action is reflected in the uncertainty principle. Whereas subsequent measurements of the same observable on the same system will have identical results, the results of measuring two complimentary observables will be uncorrelated. Back-action can be thought of as the failure for a post-measurement state to correlate with its pre-measurement state. This notion of back-action can be extended to unitary sensors. We quantify back-action as the QMI between the initial and post-interaction system state, which we term residual information $I_q^\mathcal{S,S_A}$, where
\begin{equation}\label{eq:QMI_equation_residual}
    I_q^\mathcal{S, S_A} = H\left(\rho^\mathcal{S}\right)
    + H\left(\rho^\mathcal{S_A}\right)
    - H\left(\rho^\mathcal{S,S_A}\right),
\end{equation}

\begin{figure}
    \centering
    \includegraphics[width=0.95\columnwidth]{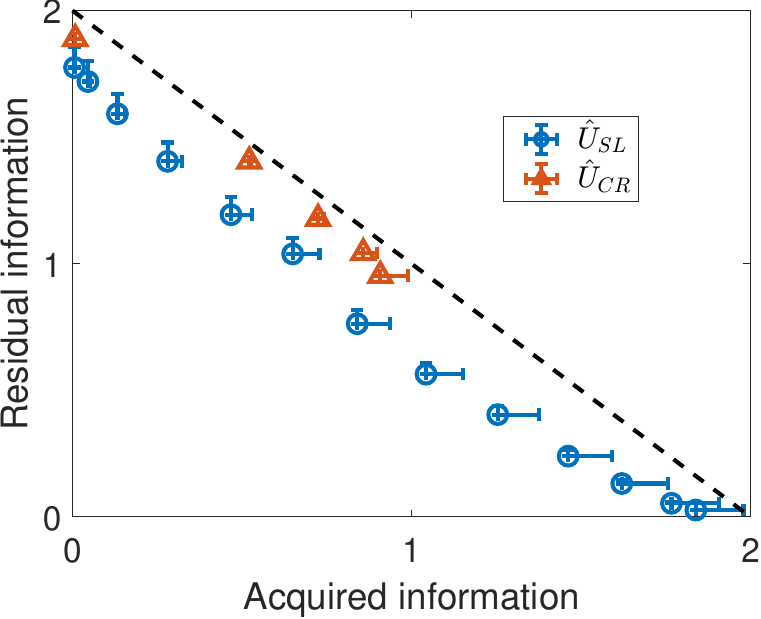}
    \caption{Experimental plot of the residual information vs acquired information (equations~\ref{eq:QMI_equation_acquired} and \ref{eq:ICmaxXY}) for $\hat{U}_{\text{CR}}$ and $\hat{U}_{\text{SL}}$ as given in equations~\ref{eq:cpauli_unitary} and~\ref{eq:SWAP_measurement_unitary}. Note that the acquired information of $\hat{U}_{\text{CR}}$ has a maximum value of 1 due to $\hat{U}_{\text{CR}}$ being a von Neumann unitary for qubits. The black dotted line reflects perfect conservation where the two information values add up to 2. Deviation of quantum mutual information values from the black dotted line is primarily due to the phase instability in the Mach-Zehnder interferometer of the respective setups of the two unitaries. The effects of these phase randomization monotonically decrease mutual information, and hence are plotted as asymmetric error bars in the plot. Errors due to imprecise waveplate rotation and off-axis LCWP rotation are not included in the error bars. Appendix \ref{secap:Errors} describes these errors in detail.}
    \label{fig:infocon}
\end{figure}

The sum of the acquired information and the residual information is conserved by \qmeasurement{}s and scales to the log of the number of dimensions in the joint system of $\mathcal{S}$ and $\mathcal{S_A}$ (short proof in appendix \ref{secap:consinfoproof}), where for qudit system states of interest of $d$ dimensions,
\begin{equation}\label{eq:info_con}
    I_q^\mathcal{S,S_A}+I_q^\mathcal{M,S_A} = 2\log[d].
\end{equation}
For non-unitary (noisy) sensors, the sum of acquired information and residual information is bounded by $2\log[d]$.
Figure \ref{fig:infocon} plots the residual vs acquired information that is reflective of equation (\ref{eq:info_con}). 
The back-action for \qmeasurements{} by quantum agents can thus be quantified by the amount of information left in the system state, with the residual information $I_q^\mathcal{S,S_A}=2$  for cases where the system state is undisturbed. In the case of bi-partite qubit interactions, the residual information is directly tied to the minimum entropy of classical information needed to perform quantum erasure \cite{Scully1982,delayedchoiceRev} of the interaction unitary. The erasure and the corresponding experimental implementation will be presented in sections \ref{sec:osdecomp} and \ref{sec:erase}.

\subsection{Specifying nature of correlation through Operator-Schmidt decomposition}\label{sec:osdecomp}
While the use of QMI gives a single-parameter quantifier of the correlations generated by a \qmeasurement{}, it does not specify the nature of these correlations.
For an arbitrary bipartite unitary, the nature of correlations generated can be found by examining the operator-Schmidt decomposition. Operator-Schmidt decomposition is a mathematical tool that is used in quantum information \cite{TysonOS_decomp,BalOS_decomp,steering_piani}.
Given a \qmeasurement{} $\hat{U}$, the operator-Schmidt decomposition is given by
\begin{equation}\label{eq:schmidt_def}
    \hat{U}=\sum_i \lambda_i\,\hat{\nu}_i^\mathcal{S}\otimes\hat{\mu}_i^\mathcal{M},
\end{equation}
where $\hat{\nu}_i^\mathcal{S}$ and $\hat{\mu}_i^\mathcal{M}$ are sets of normalized orthogonal operators such that 
\begin{equation}
\text{Tr}\left[\hat{\nu}_i^\mathcal{S}\otimes\hat{\mu}_i^\mathcal{M}\cdot\left(\hat{\nu}_i^\mathcal{S}\otimes\hat{\mu}_i^\mathcal{M}\right)^\dagger\right]=\text{dim}(\mathcal{S})\cdot\text{dim}(\mathcal{M}).
\end{equation}
The Schmidt rank is given by the number of non-zero $\lambda_i$, and the Schmidt strength is given by $-\sum p_i\log[p_i]$, with $p_i = \left|\lambda_i\right|^2$, $\sum p_i = 1$. 
The Schmidt strength corresponds to the information gain when the memory is prepared maximally entangled to a memory ancilla (see Appendix B).

Applying the operator-Schmidt decomposition to our two experimental unitaries, we find that  $\hat{U}_{\text{CR}}\left(\phi\right)$ from equation \ref{eq:cpauli_unitary} can be written in the operator-Schmidt form,
\begin{widetext}
\begin{equation}\label{eq:U2decomp}
    \hat{U}_{\text{CR}}\left(\phi\right) = 
    \cos(\phi/2)
    \left(
    \begin{array}{cc}
     \cos \left(\frac{\phi}{2}\right) & -i \sin \left(\frac{\phi}{2}\right) \\
     -i \sin \left(\frac{\phi}{2}\right) & \cos \left(\frac{\phi}{2}\right) \\
    \end{array}
    \right)^\mathcal{S}\otimes
    \left(
    \begin{array}{cc}
     1 & 0 \\
     0 & 1 \\
    \end{array}
    \right)^\mathcal{M}
    +
    \sin(\phi/2)
    \left(
    \begin{array}{cc}
     \sin \left(\frac{\phi}{2}\right) & i \cos \left(\frac{\phi}{2}\right) \\
     i \cos \left(\frac{\phi}{2}\right) & \sin \left(\frac{\phi}{2}\right) \\
    \end{array}
    \right)^\mathcal{S}\otimes
    \left(
    \begin{array}{cc}
     1 & 0 \\
     0 & -1 \\
    \end{array}
    \right)^\mathcal{M}.
\end{equation}
\end{widetext}
This unitary is of Schmidt rank 2. 
We also note that equation (\ref{eq:SWAP_measurement_unitary}) for $\hat{U}_{\text{SL}}$ is already in operator-Schmidt form, and that the unitary is of Schmidt rank 4.

For a given unitary $\hat{U}$, the possible operators $\hat{N}^\mathcal{S}$ implemented on the system are limited by the normalized linear combinations of $\lambda_i\,\hat{\nu}_i^\mathcal{S}$, where 
\begin{equation}\label{eq:operator_lin_comp}
    \hat{N}^\mathcal{S} = \sum_i c_i \lambda_i\,\hat{\nu}_i^\mathcal{S}.
\end{equation}
$\hat{N}^\mathcal{S}$ can be implemented by preparing the quantum memory in the maximally entangled state \begin{equation}
    \ket{\Phi^+}^\mathcal{M,M_A}=1/\sqrt{\text{dim}(\mathcal{M})}\sum_{i=1}^{\text{dim}(\mathcal{M})}\ket{i}^\mathcal{M}\otimes\ket{i}^\mathcal{M_A}
\end{equation}
with an ancilla ($\mathcal{M_A}$) qubit. After the \qmeasurement{} $\hat{U}$, the joint memory and ancilla state become `entangled' with operators acting on the system state:
\begin{multline}\label{eq:state_operator_equation}
    \hat{U}^\mathcal{S,M}\otimes\mathds{I}^\mathcal{M_A}\ket{\Phi^+}^\mathcal{M,M_A}\\
    = \frac{1}{\sqrt{\text{dim}(\mathcal{M})}}\sum_{i=1}^{\text{dim}(\mathcal{M})} \left[\lambda_i\ket{\mathbf{\mu_i}}^\mathcal{M,M_A}\otimes\hat{\nu}_i^\mathcal{S}\right].
\end{multline}
The projection of joint memory and ancilla state on channel dual-states \cite{choi1975} $\ket{\mathbf{\mu_i}}$ corresponds, up to a scalar, to $\lambda_i\,\hat{\nu}_i^\mathcal{S}$ acting on the system state. Alternatively, projecting the joint memory and ancilla state onto a linear combination of $\hat{\mu}_i^\mathcal{M}$ dual-states corresponds to a linear combination of operators $\lambda_i\,\hat{\nu}_i^\mathcal{S}$ acting on the system state.

From this procedure, it is immediately clear that \qmeasurements{} of lower Schmidt rank and strength are more limited in the operators $\hat{N}^\mathcal{S}$ on $\ket{\psi}$ they can sense -- the operator space spanned by the Schmidt operators of \qmeasurements{} with low Schmidt rank has limited support.
Whereas the SWAP-like \qmeasurement{} given by {$\hat{U}_{\text{SL}}(\theta=0)$} has Schmidt operators of the identity and the three Pauli operators which span the entire qubit operator space, the \qmeasurement{} given by $\hat{U}_{\text{CR}}$ only has (up to a common unitary rotation) the identity and $\hat{\sigma_x}$ as Schmidt operators. Thus a \qmeasurement{} based on the $\hat{U}_{\text{CR}}$ interaction cannot generate correlations in other bases, such as $\hat{\sigma_y}$ and $\hat{\sigma_z}$.

\FloatBarrier

\section{Quantifying back-action through erasure}\label{sec:erase}

The use of QMI to quantify residual information is imprecise about the nature of that disturbance, as is the case for using QMI for acquired information.
Certain values of $\phi$ and $\theta$ might result in very similar residual information for $\hat{U}_{\text{CR}}$ and $\hat{U}_{\text{SL}}$, but by their very natures, $\hat{U}_{\text{CR}}$ disturbance is a dephasing channel along $\hat{\sigma}_x$ and $\hat{U}_{\text{SL}}$ causes the system state to gravitate towards the initial state of the memory.

To offer a more complete description of back-action, we examine the information and operations needed to undo the interaction.
The information needed to undo a qubit unitary sensor is $2 - I_q^\mathcal{S,S_A}$, which corresponds to the number of classical bits the observer needs to send the system for the system to restore itself with local operations.
We experimentally verify this using quantum erasure \cite{Scully1982,delayedchoiceRev}, which takes in the classical bits and would undo the interaction, thus restoring the system state.

The use of the operator-Schmidt decomposition as the basis for performing erasure also serves as a demonstration of a fundamental difference between \qmeasurements{} such as  $\hat{U}_{\text{SL}}$ versus those which are von Neumann-type unitaries like $\hat{U}_{\text{CR}}$.
As the number of terms in the operator-Schmidt decomposition is equal to the number of different outcomes of the projective measurement needed to perform erasure, the operator-Schmidt rank determines the number of bits required for the erasure channel.
A TQM-based sensor for qubits, such as the von Neumann type $\hat{U}_{CR}$ sensor given here, can be erased using a single-bit channel, while a full-rank sensor, such as $\hat{U}_{SL}$, requires a two-bit channel. 

The classical bits used in erasure are generated by performing projective measurements on the memory state with an appropriate basis, where each projection result heralds a corresponding correction unitary that would return the state $\ket{\psi}$ to the system $\mathcal{S}$. Since the operator-Schmidt decomposition for a bipartite qubit unitary always admits a unitary Schmidt basis, erasure can be achieved by projecting the post-interaction memory state onto an orthonormal set of states $\hat{\mu}_i^\mathcal{M}\ket{\phi}^\mathcal{M}$, with the corresponding erasure unitary to restore the system state being $\hat{\nu}^{\dagger\mathcal{S}}$. 

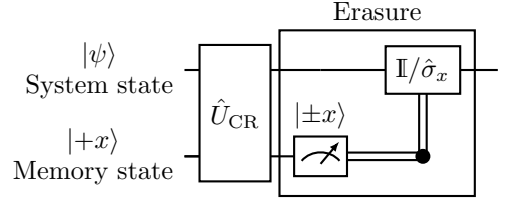
\begin{figure}
    \centering
    \begin{quantikz}
        \lstick{$\ket{\psi}$\\System state} & [-0.3cm]\gate[wires=2]{\hat{U}_{\text{CR}}} & [-0.2cm]\qw\gategroup[wires=2,steps=2,style={inner sep=2pt}]{Erasure} & \gate{\{\hat{\nu}_1^{\dagger},\hat{\nu}_2^{\dagger}\}} & \qw \\[-0.4cm]
        \lstick{$\ket{+x}$\\Memory state} & \qw & \meter{\ket{\pm x}} & \cwbend{-1} & 
    \end{quantikz}
    \caption{Circuit diagram for the erasure of the \qmeasurement{} $\hat{U}_{\text{CR}}$. After the \qmeasurement{}, the memory state is projected onto the $\ket{\pm x}\!\!\bra{\pm x}$ states. Depending on the result of this projection, either $\hat{\nu}_1^{\dagger}$ or $\hat{\nu}_2^{\dagger}$} is applied.
    \label{fig:cpauli_erase_circuit}
\end{figure}

\begin{figure*}
    \centering
    \includegraphics[width=0.45\textwidth]{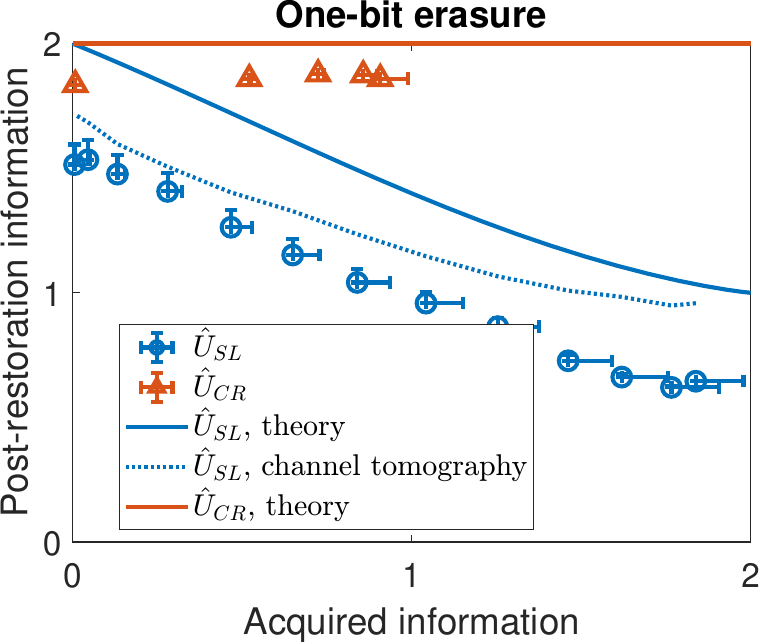}
    \includegraphics[width=0.45\textwidth]{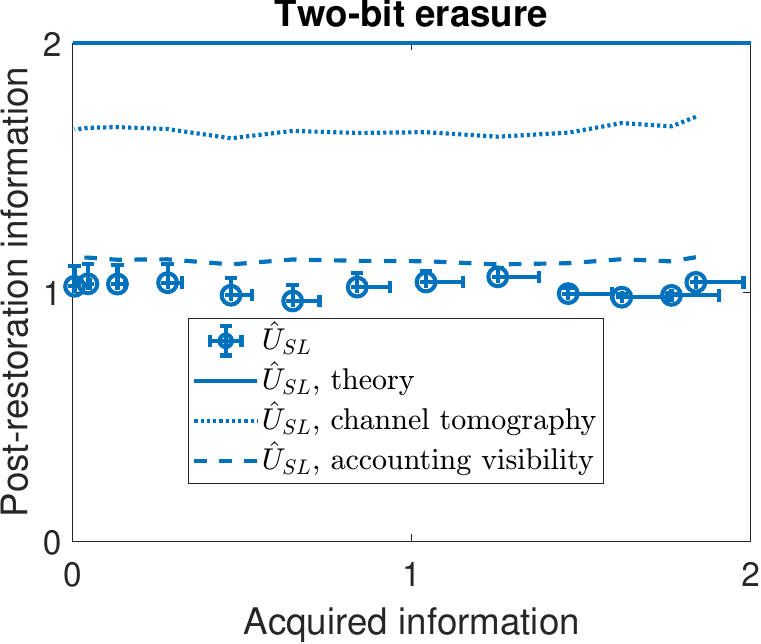}
    \caption{Quantum erasure of two $\hat{U}_{CR}$ and $\hat{U}_{SL}$ (equations \ref{eq:cpauli_unitary} and \ref{eq:SWAP_measurement_unitary}, respectively) by using classical information to restore the original state $|\psi\rangle$.
    The left plot shows that with a one-bit channel, the post-restoration information (bits) is constant for $\hat{U}_{CR}$ but decreases for $\hat{U}_{SL}$ with respect to the acquired information (bits).
    One bit is sufficient to erase a traditional quantum measurement like UCR, but not the more general interaction unitaries like USL. 
    The right plot shows that when a two-bit channel is used for $\hat{U}_{SL}$, its post-restoration information becomes constant, as its complete erasure fundamentally requires two bits due to the Schmidt rank of $\hat{U}_{SL}$ being $2^2$. While a perfect two-bit restoration would yield a value of 2 (solid line), our experimental result was only $1.01\pm0.05$ (markers). This value is lower than the result in the left plot because the two-bit measurement procedure was degraded by a limited two-photon visibility of $\sim0.78$, an experimental imperfection that did not affect the one-bit erasure scheme, which did not use two-photon interference.} 
    
    \label{fig:erase_QMI_U4}
\end{figure*}

In our experiment, the application of the correction unitary is carried out by changing the basis states used in tomography.
For $\hat{U}_{\text{CR}}$, the two Schmidt operators for the memory are the identity and $\hat{\sigma}_z$, which when acting on the initial memory state $\ket{+x}$ and unknown system state $\ket{\psi}$ results in
\begin{multline}
    \hat{U}_{\text{CR}}^\mathcal{S,M}\left(\phi\right) \, \ket{\psi}^\mathcal{S}\otimes\ket{+x}^\mathcal{M}\\
    = \frac{1}{\sqrt{2}} [\cos\left(\phi/2\right)\hat{\nu}_1^\mathcal{S}\ket{\psi}^\mathcal{S}\otimes\ket{+x}^\mathcal{M} \\
    + \sin\left(\phi/2\right)\hat{\nu}_2^\mathcal{S}\ket{\psi}^\mathcal{S}\otimes\ket{-x}^\mathcal{M}],
\end{multline}
where $\hat{\nu}_1$ and $\hat{\nu}_2$ are given in equation (\ref{eq:U2decomp}).
Therefore, erasure involves projecting the memory state onto the $\ket{\pm x}\!\!\bra{\pm x}$ states and subsequently performing either $\hat{\nu}_1^{\dagger}$ if the result is $\ket{+x}$  or $\hat{\nu}_2^{\dagger}$ if the result is $\ket{-x}$ on the system state, a procedure similar to that in quantum eraser experiments.
The corresponding circuit diagram for the erasure of $\hat{U}_{\text{CR}}$ can be found in figure \ref{fig:cpauli_erase_circuit}, with the resulting \textit{post-restoration information}, the QMI {(equation~\ref{eq:QMI_equation_residual})} between the system ancilla ($\mathcal{S_A}$) and the system ($\mathcal{S}$) after erasure, found in figure \ref{fig:erase_QMI_U4}.

The application of $\hat{U}_{\text{SL}}$ on the initial memory state $\ket{\phi}$ and unknown system state $\ket{\psi}$ results in
\begin{align}
    &\hat{U}_{\text{SL}}^\mathcal{S,M}\ket{\psi}^\mathcal{S}\otimes\ket{\phi}^\mathcal{M}\\
    &= \,\frac{1}{2}(
    -(1+\sin (\theta{-\pi/2}))\,\ket{\psi}^\mathcal{S}\otimes\ket{\phi}^\mathcal{M}\\
        &\quad -[1-\sin (\theta{-\pi/2}))\,\hat{\sigma}_z^\mathcal{S}\ket{\psi}^\mathcal{S}\otimes\hat{\sigma}_z^\mathcal{M} \ket{\phi}^\mathcal{M}\\
        &\quad+i\,\cos (\theta{-\pi/2}) \hat{\sigma}_x^\mathcal{S}\ket{\psi}^\mathcal{S}\otimes\hat{\sigma}_y^\mathcal{M} \ket{\phi}^\mathcal{M}\\
        &\quad-i \cos (\theta{-\pi/2}) \hat{\sigma}_y^\mathcal{S}\ket{\psi}^\mathcal{S}\otimes\hat{\sigma}_x^\mathcal{M} \ket{\phi}^\mathcal{M}] .
\end{align}

For $\hat{U}_{\text{SL}}$, the four Schmidt operators $\hat{\mu}^\mathcal{M}$ are the identity and the three Pauli operators, and thus the set of states given by $\hat{\mu}^\mathcal{M}\ket{\phi}$ where $\ket{\phi}=\ket{+z}$ is not mutually orthogonal. After all, no channel with a qubit output can give a set of four mutually orthogonal outcomes. 
Instead, the erasure of $\hat{U}_{\text{SL}}$ requires the inclusion of a memory qubit ancilla that is maximally entangled with the initial memory qubit in the joint state given by $\ket{\Phi^+}^\mathcal{M,M_A}$. 
The application of the four Schmidt operators $\hat{\mu}^\mathcal{M}$ on $\ket{\Phi^+}^\mathcal{M,M_A}$ results in the four canonical Bell states. 

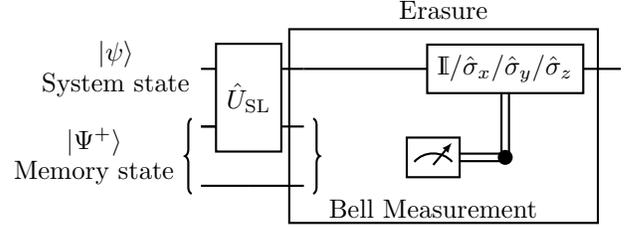
\begin{figure}
    \centering
    \begin{quantikz}
        \lstick{$\ket{\psi}$\\System state} & [-0.3cm]\gate[wires=2]{\hat{U}_{\text{SL}}} & [-0.2cm]\qw\gategroup[wires=5,steps=3,style={inner sep=2pt}]{Erasure} & [-0.4cm] \qw & [-2.2cm] \gate{\{\mathds{I},\hat{\sigma}_x,\hat{\sigma}_y,\hat{\sigma}_z\}} & \qw\\[-0.4cm]
        \lstick[wires=3]{$\ket{\Psi^{+}}$\\Memory state} & \qw & \qw \rstick[wires=3]{} &  & & \\[-0.7cm]
         & & & |[meter]| & \cwbend{-2} \\[-0.4cm]
         & \qw & \qw &  & & \\[-0.2cm]
         & & & \midstick{Bell Measurement} & & 
    \end{quantikz}
    \caption{Circuit diagram for the erasure of the \qmeasurement{} $\hat{U}_{\text{SL}}$. After the \qmeasurement{}, the memory state is projected onto the canonical Bell states. Depending on the result of this projection, either the identity or one of the Pauli operators is applied. The erasure procedure can be considered a quantum teleportation procedure for $\hat{U}_{\text{SL}}$.}
    \label{fig:swap_erase_circuit}
\end{figure}

Erasure of $\hat{U}_{\text{SL}}$  can thus be performed using a 2-bit channel, by projecting the joint memory and memory ancilla state onto the Bell basis and applying the corresponding $\hat{\nu}^{{\dagger},{\mathcal{S}}}$ unitary on the system state. This process is akin to a quantum teleportation protocol \cite{teleport}, and the corresponding circuit diagram for the erasure of $\hat{U}_{\text{SL}}$ can be found in figure \ref{fig:swap_erase_circuit}.

We compare with erasure relying on a single-bit channel,  emulating a 1-bit classical memory ancilla by projecting the joint memory and memory ancilla states incoherently onto the correlated ($\ket{+z,+z}\!\!\bra{+z,+z}$, $\ket{-z,-z}\!\!\bra{-z,-z}$) and anti-correlated ($\ket{+z,-z}\!\!\bra{+z,-z}$, $\ket{-z,+z}\!\!\bra{-z,+z}$) states, and apply the $\hat{\sigma}_x$ or $\hat{\sigma}_z$ unitary as correction correspondingly.
 
Figure \ref{fig:erase_QMI_U4} shows the post-restoration information after the application of correction unitary $\hat{\nu}^{{\dagger}{\mathcal{S}}}$ on the system, for one-bit (left) and two-bit (right) classical erasure channels.
Ideally, the 1-bit channel permits perfect erasure of $\hat{U}_{CR}$,  while a 2-bit channel is required for erasure of $\hat{U}_{SL}$.  Indeed, with a 1-bit channel, we see nearly perfect restoration of information for $\hat{U}_{CR}$, regardless of measurement strength -- on the other hand, for $\hat{U}_{SL}$ we see that the effectiveness of the erasure drops with increasing interaction strength.  By contrast, for 2-bit erasure of $\hat{U}_{SL}$ (right), we see that the restored information remains essentially constant regardless of measurement strength.  Although our low two-photon visibility ($\sim 0.78$) reduced the fidelity of Bell-state measurements and prevented us from accomplishing perfect erasure, we take this constancy as an indication that the two-bit channel is in-principle sufficient to erase swap-like measurements of arbitrary strength.  These results support our assertion that the channel capacity required to erase the effects of a measurement is determined by the Schmidt rank of the measurement unitary.





\begin{figure}
    \centering
    \includegraphics[width=0.95\columnwidth]{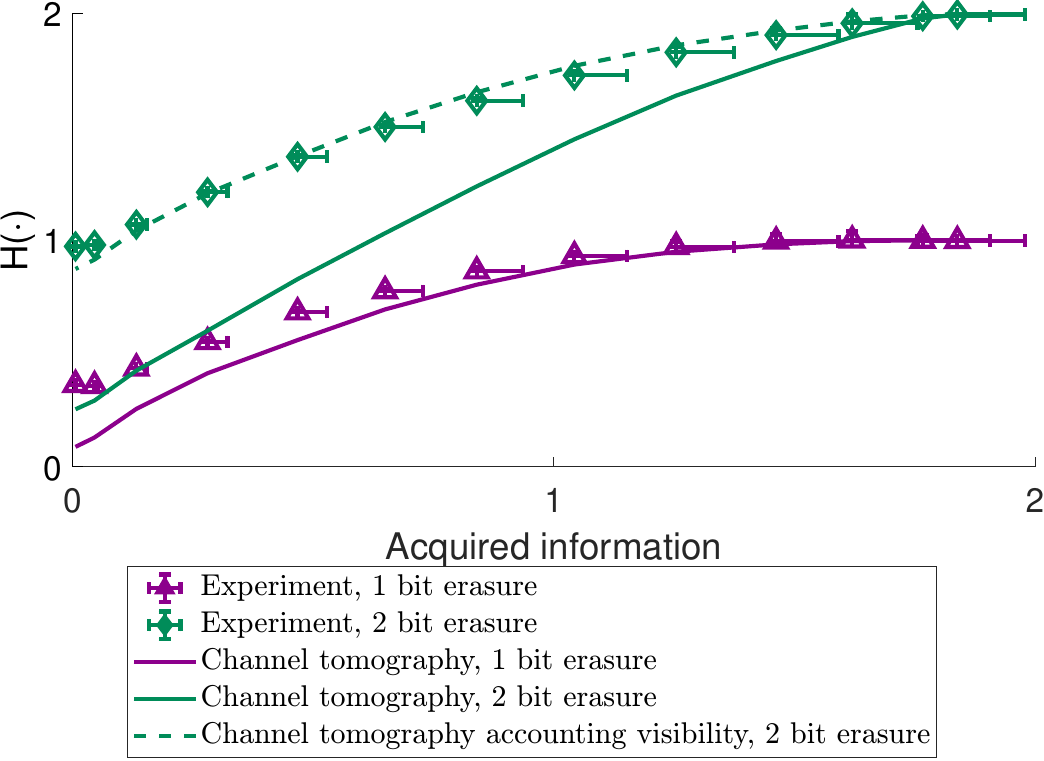}
    \caption{The need for the use of 2 bits of information to perform erasure for $\hat{U}_{\text{SL}}$ (equation \ref{eq:SWAP_measurement_unitary}) exists even when $\hat{U}_{\text{SL}}$ is tuned close to the identity and the entropy of the two bits is much less than one. Here, we plot the Shannon entropy, $H$, of the bits used to perform erasure for $\hat{U}_{\text{SL}}$. As we have shown in figure \ref{fig:erase_QMI_U4}, $\hat{U}_{\text{SL}}$ can only be erased with 2 bits of information unless it is tuned to identity. In the above plot, a region that can be inferred from channel tomography where those 2 bits have an entropy of less than one exists. This result is obscured in our erasure experiment due to insufficient two-photon visibility.}
    \label{fig:erase_bits}
\end{figure}

A regime also exists at low interaction strengths of $\hat{U}_{\text{SL}}$ where the number of bits needed to perform perfect erasure is 2, but the entropy of those bits is less than 1. Figure \ref{fig:erase_bits} shows the entropy of the bits used to perform erasure for $\hat{U}_{\text{SL}}$.

The application of the Schmidt-operator decomposition described in section \ref{sec:osdecomp} to find erasure procedures also warrants caution when used for bipartite unitary of higher dimensions. While our assertion that the possible operators implementable for a certain unitary is given by the Schmidt-operator basis still holds (equation (\ref{eq:operator_lin_comp})), the correspondence between the maximum value of the acquired information and Schmidt-strength no longer applies. An example is the controlled-SWAP discussed in appendix \ref{secap:QMI_qubit_unitaries}.

\section{Conclusion}

In this work, we extended the concept of quantum measurement by classical observers to a broader class of ``quantum sensors'' given by general unitaries which may go beyond the standard von Neumann--style $\hat{U}_{\text{CR}}$ to include, for instance, a SWAP-like unitary $\hat{U}_{\text{SL}}$.  While the latter operation can capture more information about a system (in principle, {\em all} of it), we show that it also leads to more disturbance.  Heisenberg already taught us that a classical observer measuring the position $X$ perfectly must disturb the momentum $P$; a {\em quantum} observer may ``sense'' both $X$ {\em and} $P$ perfectly, but in so doing they would also disturb both, effectively destroying the original state. 
Erasing such total disturbance would require quantum teleportation, which is well known for a qubit system to require 2 classical bits.
We show that even weak SWAP-like interactions require 2 bits of information to erase. 
We demonstrate this constraint experimentally, in the case of a partial-SWAP unitary ($\hat{U}_{\text{SL}}$), confirming that a second bit of channel capacity was required even when the entropy of the classical bits used in restoring the observed system was less than one.


Our work highlights the qualitative and quantitative distinctions between sensors for quantum and classical agents. We introduced analysis techniques for these sensors for quantum agents which we believe will prove powerful as technology progresses to permit larger and larger systems to be placed under fully quantum control. Our work grounds foundational questions about quantum agency in the operational and experimental framework of quantum sensing and erasure.


\section{Acknowledgments}
This work was supported by the Foundational Questions Institute under FQXiRFP-1819, the John Templeton Foundation under grant ID 63209, and NSERC under Discovery Grant RGPIN-2020-05767.  Additional support came from the Fetzer Franklin Fund of the John E. Fetzer Memorial Trust and the QuEnSi quantum alliance, under NSERC ALLRP 578468 - 22. A. M. S is a fellow of CIFAR.

\FloatBarrier

\bibliography{references}

\appendix

\section{Conservation of acquired and residual information}\label{secap:consinfoproof}
We provide a simple proof of equation (\ref{eq:info_con}) in this section. We begin by examining the sum of the acquired and residual information in terms of entropic quantities, where
\begin{align}
\begin{split}
    I_q^\mathcal{S,S_A}+&I_q^\mathcal{M,S_A} =\\
    &H\left(\rho^\mathcal{S}\right)
    + H\left(\rho^\mathcal{S_A}\right)
    - H\left(\rho^\mathcal{S,S_A}\right)\\
    &+ H\left(\rho^\mathcal{S_A}\right)
    + H\left(\rho^\mathcal{M}\right)
    - H\left(\rho^\mathcal{M,S_A}\right).
\end{split}
\end{align}
Next, we note that since the initial quantum state in the joint space of $\mathcal{S}$, $\mathcal{S_A}$ and $\mathcal{M}$ is a pure state and the evolution is unitary, the post-evolution state $\rho^\mathcal{S,S_A,M}$ is also pure. Since $\rho^\mathcal{S,S_A,M}$ is pure, we can also conclude that any bipartition of the state $\rho^\mathcal{S,S_A,M}$ will result in the two partitions having equal entropies, and as such $H\left(\rho^\mathcal{S}\right) = H\left(\rho^\mathcal{M,S_A}\right)$ and $H\left(\rho^\mathcal{M}\right) = H\left(\rho^\mathcal{S,S_A}\right)$. Additionally, since the state in $\mathcal{S,S_A}$ initially a maximally entangled state and that $\mathcal{S_A}$ does not interact with the unitary, the marginal state of $\rho^\mathcal{S_A}$ is a maximally mixed state, with $H\left(\rho^\mathcal{S_A}\right) = \log_2\left[\text{dim}(\mathcal{S_A})\right]$. As a result, we conclude that
\begin{equation}
    I_q^\mathcal{S,S_A}+I_q^\mathcal{M,S_A} = 2\cdot H\left(\rho^\mathcal{S_A}\right) = 2\cdot\log_2\left[\text{dim}(\mathcal{S_A})\right].
\end{equation}

\section{Maximum QMI for bipartite qubit-unitaries}\label{secap:QMI_qubit_unitaries}
To find the QMI between the initial system state and the post-interaction memory state, we employ a system ancilla that is initially maximally entangled with the system state, such that
\begin{equation}\label{eq:max_ent_app}
     \ket{\Phi^+}^\mathcal{S,S_A}=1/\sqrt{2}\left(\ket{00}^\mathcal{S,S_A}+\ket{11}^\mathcal{S,S_A}\right).
\end{equation}
The QMI of the memory and the initial system state is given by $I^{\mathcal{S,S_A}}$.

The equation for QMI between the memory and system state can be rewritten in the form
\begin{align}
      I_q^\mathcal{S_A, M, M_A} &= H\left(\rho^\mathcal{M, M_A}\right)
    - H\left(\rho^\mathcal{S_A,M, M_A}|\rho^\mathcal{S_A}\right)\\
    &= H\left(\rho^\mathcal{M, M_A}\right)
    - H\left(\rho^\mathcal{S_A,M, M_A}\right) + H\left(\rho^\mathcal{S_A}\right)\\
    &= H\left(\rho^\mathcal{M, M_A}\right)
    - H\left(\rho^\mathcal{S}\right) + H\left(\rho^\mathcal{S_A}\right)
\end{align}
where we have introduced the memory ancilla $\mathcal{M_A}$ for generality. 
For a maximally entangled system state $|\phi^+\rangle^\mathcal{S,S_A}$, $H(\rho^\mathcal{S})=H(\rho^\mathcal{S_A})$. And thus, $I_q^\mathcal{S_A,M,M_A}=H(\rho^\mathcal{M,M_A})$.

Additionally, application of the unitary to the state given in \ref{eq:max_ent_app} and an arbitrary pure initial memory state $\ket{\gamma}^\mathcal{M,M_A}$ results in
\begin{multline}\label{eq:operator_state}
|\psi\rangle =\mathds{I}^\mathcal{S_A,M_A}\otimes\hat{U}^\mathcal{S,M}\,\ket{\Phi^+}^\mathcal{S,S_A}\otimes\ket{\gamma}^\mathcal{M,M_A}\\
= \sum\lambda_i\ket{\mathbf{\nu_i}}^\mathcal{S,S_A}\otimes\left(\hat{\mu}_i^\mathcal{M}\otimes\mathds{I}^\mathcal{M_A}\ket{\gamma}^\mathcal{M,M_A}\right).
\end{multline}
The structure of equation (\ref{eq:operator_state}) is similar to that of equation (\ref{eq:state_operator_equation}). 
For bipartite-qubit unitaries, there is always a Schmidt-operator decomposition where $\hat{\mu}_i$ and $\hat{\nu}_i$ are unitaries \cite{Watrous}. This implies that $\ket{\mathbf{\nu_i}}^\mathcal{S,S_A}\otimes\left(\hat{\mu}_i^\mathcal{M}\otimes\mathds{I}^\mathcal{M_A}\ket{\gamma}^\mathcal{M,M_A}\right)$ is a normalized state. Here, $\ket{\mathbf{\nu_i}}^\mathcal{S,S_A}$ is a set of orthogonal states, and by setting $|\gamma\rangle^{\mathcal{M,M_A}}=|\Phi^+\rangle^{\mathcal{M,M_A}}$, the composite memory and memory ancilla state $|\xi_i\rangle=\left(\hat{\mu}_i^\mathcal{M}\otimes\mathds{I}^\mathcal{M_A}\ket{\gamma}^\mathcal{M,M_A}\right)$ will also be a set of orthogonal states. 
The joint memory state $\rho^\mathcal{M,M_A}$ can be obtained by tracing the state $|\psi\rangle$ (\ref{eq:operator_state}) across $\mathcal{S}$ and $\mathcal{S_A}$:
\begin{align}
   \rho^\mathcal{M_A,M}= \mathrm{Tr}_\mathcal{S,S_A}(|\psi\rangle\langle\psi|) = -\sum_{i}|\lambda_i|^2  |\xi_i\rangle\langle\xi_i|
\end{align}
Having $\ket{\Phi^+}^\mathcal{M,M_A}$ as the initial memory state results in the entropy $H\left(\rho^\mathcal{M_A,M}\right) = -\sum_i|\lambda_i|^2\log[|\lambda_i|^2]$, and hence the acquired information is equal to the Schmidt strength
\begin{equation}\label{eq:QMI_eq_schmidt}
    I_q^\mathcal{S_A,M,M_A} = -\sum|\lambda_i|^2\log[|\lambda_i|^2]
\end{equation}
For highly nonlocal gates such as CNOT and SWAP, this Schmidt strength is maximized and thus the acquired information is maximized.

Equation (\ref{eq:QMI_eq_schmidt}) exclusively applies to the bipartite unitaries that admit a Schmidt-operator decomposition where the set of operators $\hat{\mu}_i^\mathcal{M}$ is unitary, and as such, always applies to bipartite qubit-unitaries. The QMI does not equal the Schmidt strength in general for higher dimensions, for example, the ququad-qubit bipartite unitary of controlled-swap
\begin{equation}
    \ket{0}\!\!\bra{0}^{\mathcal{M}1}\otimes\hat{U}_{swap}^{\mathcal{M}2,\mathcal{S}}+\ket{1}\!\!\bra{1}^{\mathcal{M}1}\otimes\mathds{I}^{\mathcal{M}2,\mathcal{S}}
\end{equation}
has a Schmidt strength of $\sim1.86$, whereas the QMI between the memory and the system is 2 if ${\mathcal{M}1}$ is prepared in the $\ket{0}^{\mathcal{M}1}$ state.

\section{Systematic Errors}\label{secap:Errors}
The $\hat{U}_{\text{CR}}$ setup is adversely affected by four sources of systematic error. The first is due to phase fluctuation in our Mach-Zehnder interferometer of about $0.3$ rad. The second is the calibration error in the polarization rotation axis of the LCWPs of about $0.06$ rad, and this error, in particular, has the effect of reducing $I_{c,max}$ for a corresponding value of the acquired information as plotted in figure \ref{fig:QMI_vs_CMI}. The third is the waveplate angle uncertainty for both the state preparation and tomography stages of the experiment of about $0.01$ rad. Finally, there is a polarization-dependent loss of at most $\sim 7\%$ present in the setup.

The  $\hat{U}_{\text{SL}}$ setup is primarily limited by two factors. Firstly, spatial misalignment of the interferometer paths leads to polarization-dependent coupling at the output/measurement couplers. The setup has 3 interferometers, each with two paths, making the total possible number of paths for a photon to traverse the setup to be 8. Between the 8 paths, the best and worst coupled paths have a coupling difference of $\sim24\%$. Secondly, the Mach-Zehnder phase is manually locked, with phase drifts of up to 0.08 $rad$.

All error bars plotted in this work represent systematic errors caused by phase randomization of our interferometer. As mutual information only decreases with added randomness, all error bars are thus asymmetric. 

\section{Schmidt rank and bits required for erasure of bipartite qubit unitaries}\label{secap:erasurebits}
For any bipartite qubit unitary interactions, there is always a Schmidt decomposition such that the set of operators $\hat{\nu}_i^\mathcal{S}$ given in equation (\ref{eq:state_operator_equation}) are unitaries\cite{BalOS_decomp}. Therefore, to undo the bipartite unitary, the memory and the memory ancilla needs to be projected onto the set of states $\ket{\mathbf{\mu_i}}^\mathcal{M,M_A}$. The result of said projection would put the system into the $\hat{\nu}_i^\mathcal{S}\ket{\psi}^\mathcal{S}$ state, with $\ket{\psi}^\mathcal{S}$ being the initial system state. To restore the system state, the unitary $\hat{\nu}_i^{\dagger\mathcal{S}}$ must be applied to the system. As the number of outcomes of the  memory state projection is equal to the Schmidt rank, the number of bits needed to perform erasure must be greater or equal to $\log_2$ of the Schmidt rank.

\end{document}